# Electronic Polarizability Induced Cooper-like Pairing and Energy Gap in High-$T_c$ superconductors

Yizhak Yacoby[1*], Davide Ceresoli[2], Livia Giordano[3,4,5] and Yang Shao-Horn[3,4,6,7*]

1. Racah Institute of physics, Hebrew University, 91904, Jerusalem, Israel
2. Consiglio Nazionale delle Ricerche, Istituto di Scienze e Tecnologie Chimiche "G. Natta" (CNR-SCITEC), 20133 Milan, Italy
3. Electrochemical Energy Laboratory, Massachusetts Institute of Technology, Cambridge MA 02139, USA
4. Research Laboratory of Electronics, Massachusetts Institute of Technology, Cambridge MA 02139, USA
5. Dipartimento di Scienza dei Materiali, Università di Milano-Bicocca, 20125 Milan, Italy
6. Department of Mechanical Engineering, Massachusetts Institute of Technology, Cambridge MA 02139, USA
7. Department of Materials Science and Engineering, Massachusetts Institute of Technology, Cambridge MA 02139, USA

**Corresponding Authors**

Email: shaohorn@mit.edu, yizhak.yacoby@mail.huji.ac.il

**Abstract**

High temperature superconducting materials have been known since the pioneering work of Bednorz and Mueller in 1986. While the microscopic mechanism responsible for high Tc superconductivity is still debated, most materials showing high Tc contain highly electronic polarizable ions, suggesting that the mechanism driving high Tc superconductivity can be related to the ion electronic polarizability in high Tc materials. Here we show that a free charge carrier polarizes the ions surrounding it and the total electrical potential generated by the charge carrier itself and the polarized ions becomes attractive in some regions of space. Our results on bulk FeSe, monolayer FeSe on $SrTiO_3$ and $La_2CuO_4$ are in excellent agreement with the experiments. The fact that the electronic polarizability explains correctly and quantitatively the superconductivity parameters: Tc, gap and paring energies of both pnictides and cuprates with similar polarizability parameters, suggests that the same model may be applicable to other material systems within these groups as well as other high Tc groups.

**Introduction**

Since the discovery of superconductivity in cuprates [1] intense research has focused on increasing the critical temperature ($T_c$) of new classes of superconducting materials. This intense research has led to a steady increase of $T_c$ in cuprates, which surpass the technologically relevant threshold of liquid nitrogen with YBaCuO in 1987 ($T_C$ = 93 K) [2]. More recently, new classes of materials have been shown to have promisingly high $T_c$, which include bismuthates (i.e. $Ba_{1-x}K_xBiO_3$) [3] and pnictides [4]. Recent experiments at high pressure have found hydrogen sulfide to be superconducting at 203 K and 150 GPa [5]. Even more recently, carbonaceous Sulphur hydrides have been reported as the first room-temperature superconductor with a $T_c$ of 288 K at 267 GPa [6].

Despite the intense research in this field, the underlying mechanism responsible for superconductivity in high $T_c$ superconductors is not understood [7]. Electron pairing, i.e. the formation of Cooper or Cooper-like pairs [8] is responsible for the transition to the superconducting phase (Bardeen Cooper Schrieffer BCS theory), where paring is induced by bosonic excitations of lattice or electronic subsystems or both. Bosonic excitations reported to contribute to superconductivity are phonons [9, 10] in the BCS theory, or plasmons, and antiferromagnetic spin fluctuations in BCS-like models [11, 12]. Lattice or spin polarons may also have a role in case of strong coupling, where the polaron-polaron interaction can cause the formation of bi-polarons leading to superconductivity in Hubbard-like models [13, 14].

We note that all the superconducting materials mentioned above contain highly electronic polarizable ions. The polarizability of group-16 anions increases along the series ($O^{2-}$: 3.06-4.36, $S^{2-}$: 9.57-11.35, $Se^{2-}$: 11.81-12.46, in MKS units of $10^{-40}$ $Cm^2/V$ [15]). Likewise, the polarizability of cations of the sixth period increases along the groups: ($Ba^{2+}$: 1.72-2.78 [15], $La^{3+}$: 2.80-4.19 [16], $Tl^+$: 5.78 [15], $As^{5+}$: 1.78 [16], in units of $10^{-40}$ $Cm^2/V$). This realization suggests that ion electronic polarizability can have a role in the electron or hole pairing. While electron-only, non-phonon mechanisms of superconductivity have been suggested in early studies by Little [17], Ginzburg [18], Kohn and Luttinger [19], only a few studies have explored the role of ionic polarization on electron pairing [20].

Bussmann-Holder and coworkers [21] have pointed out a possible link between ferroelectricity and superconductivity in oxides due to instability of the $O^{2-}$ ion, giving rise to a dynamical change in the *p-d* hybridization at optical phonon frequencies. A similar mechanism has been proposed by Callaway and coworkers [22], where quadratic terms representing the interactions among the Cu ions and the polarization fluctuations of the oxygen ions are added to the Hubbard Hamiltonian describing strong correlations of electrons in cuprates. Unfortunately, their Hubbard model can be solved only by exact-diagonalization for small $CuO_4$ clusters, and an accurate estimate of $T_c$ with this method is computationally too expensive. Recently, Atwal and coworkers [23] have reported that electron-electron attraction can be induced by dynamic electron correlation, resulting in polarization waves that promote electron pairing. Berciu and collaborators [20] have shown the formation of electronic polarons, induced by the presence of highly polarizable $As^{3-}$ anions, and the formation of bi-polaron pairs in Fe-based pnictides, which can have a role in the electron pairing in these materials. However, their model has neglected the electron-electron repulsion, making it difficult to quantitatively estimate the magnitude of this electronic effect. More recently, based on angle resolved photo electron emission spectroscopy (ARPES) measurements and electron energy loss spectroscopy (EELS) measurements, Song et. al. [24] find that the electron phonon interaction (EPI) coefficient and the superconductivity energy gap (SEG) are linearly related SEG=SEG0+B*EPI, where SEG0 =9.5 meV, indicating that the energy gap is still large even when EPI=0. Consequently Song el.al [24] conclude that another mechanism accounts for most of the superconductivity gap energy. In addition, they discard, on various grounds, models based on strain and on high polarizability due to soft vibration modes [24].

From a pure ab initio perspective, density functional theory for superconductors (SCDFT) is one of the first-principles methods to compute $T_c$, and it can treat the electron-phonon interaction, electronic Coulomb interaction, and spin fluctuation (SF) fully non-empirically. This method has been applied mainly to the electron-phonon coupled superconductors [25] and its predictive power has been improved recently by the introduction of refined functionals [26]. A pure ab initio theory is however difficult to interpret in terms of simple models that focus on one particular coupling mechanism, such as electron phonon interaction EPI.

Here we propose a theoretical framework that describes the interaction between two charge carriers in presence of polarizable ions. The principles of the theory are as follows: We consider two free charge carriers (energies above Fermi level, holes for simplicity). We shall refer to one as the probe and the second as charge carrier. Both the probe and charge carrier polarize directly and indirectly their surrounding ions. The electrical potential induced by the probe and the polarized ions can be positive (repulsive) or negative (attractive) in different regions in space. If the probe and charge carrier are far from each other they will not feel the potential induced by the other. But if they are close they can feel the potential induced by the other and form a two-charge carrier wavefunction. The two charge carrier wave function is a quantum state. If the corresponding energy is negative relative to the Fermi level and the charge carriers have opposite spin the two charge carriers will be paired into a singlet state. This analysis builds upon the previous work from one of us [27], which we have extended from a simplistic artificial material system and plane wave functions to the structures of real systems and proper solutions of the Schrodinger equation.

We use Density Functional Theory to calculate the one-electron Bloch wavefunctions. In the presence of two charge carriers, each will be surrounded by the same but displaced potential map. If the distance between the charge carrier and the probe is small enough, each charge carrier feels the potential induced by the other. We now treat the potential function (positive and negative regions) as a perturbation and solve for the two charge-carrier wavefunction. To solve the Schrödinger equation, we expand this wavefunction in terms of products of two one charge-carrier wave functions, calculate the corresponding matrix elements and solve the Schrödinger equation. There are a number of solutions, where the one with the lowest energy, if negative, is the paring energy. In addition, we obtain an upper limit to the energy of the BCS multi-electron wavefunction. If the energy is negative, we define it here as minus the gap energy. By making some justified assumptions: (i) the interaction between the charge carries and the ion electrons is treated electrostatically due to the limited size of the region of interest (<1 nm), (ii) Since the electrons mass is very small the electronic polarization response is treated as instantaneous; and (iii) All ions close to the probe are involved; we computed the pairing energy and superconducting energy gap, for several systems based on FeSe and $La_2CuO_4$.

Our quantitative results demonstrate that the pairing of charge carriers induced by electronic polarization of anions can account for the pairing energies. The results show that the superconductive properties (namely, paring and gap energies) of bulk and one-unit-cell FeSe on STO, and $La_2CuO_4$ can be understood quantitatively. In particular, the results show that in the systems we investigated the polarizabilities needed to produce the known superconducting gap and pairing energies is within the range of electronic polarizability values reported in the literature [28] [15]. These results can pave the way for a deeper understanding of high-$T_C$ superconducting mechanism, and can be leveraged for the design of novel materials with elevated superconducting temperatures.

**Theory**

High $T_c$ superconductors possess ions with large electronic polarizabilities. We hypothesize that under proper conditions these polarizabilities may lead to electron pairing and superconductivity. We computed the polarization induced directly by a free charge carrier, and indirectly by the surrounding polarized ions self consistently. We used for example holes as charge carriers and we included screening in our approach. We further calculated the electrical potential as a function of probe position including the repulsive potential of the probe. Positive and negative potentials mean repulsive and attractive potentials between holes, respectively and the corresponding relation holds for electrons.

We consider the following one-charge-carrier Hamiltonian:

$$H_a = \frac{p_1^2}{2m} + e\emptyset_0(\vec{r}_1) + e\emptyset_p(\vec{r}_1,\vec{r}_1) \qquad (1)$$

Here $r_1$ is the probe position. $\emptyset_p(\vec{r}_1,\vec{r}_2)$ is the potential at $\vec{r}_2$ induced indirectly by the polarized surrounding ions. $\emptyset_0(\vec{r}_1)$ is the potential due to all other charges. The ionic polarization is the product of the polarizability constant and the total electric field at the ion center. This electric field at one ion is the sum of the partially screened electric field of the probe and the electric field generated by the other polarized ions calculated self consistently, so that the total electric field and the total potential are proportional to the probe charge. The one-charge-carrier Schrodinger equation is:

$$H_a\Psi_{\vec{K}}(\vec{r}_1) = \varepsilon_{\vec{K}}\Psi_{\vec{K}}(\vec{r}_1) \qquad (2)$$

Where $\varepsilon_{\vec{K}} = \xi_{\vec{K}} + E_F$, $E_F$ is the Fermi energy and $\Psi_{\vec{K}}(\vec{r}_1)$ is the Bloch wavefunction

$$\Psi_{\vec{K}}(\vec{r}_1) = \varphi_{\vec{K}}(\vec{r}_1)\exp(i\vec{K}\cdot\vec{r}_1) \qquad (3)$$

We evaluated the Bloch wavefunctions of select systems based on FeSe and $La_2CuO_4$ using Density Functional Theory (DFT) calculations (Details of the DFT calculations are in the Supporting Information), where the effect of polarization of the surrounding ions on the probe is included implicitly at the mean field level.

The FeSe-SrTiO$_3$ was modeled as a symmetric slab with 7 SrO/TiO$_2$ layers, where the FeSe monolayer was deposited on the TiO$_2$-terminated surfaces, with the Se atoms on top of Ti atoms. In addition, we chose the low-temperature-tetragonal (LTT) polymorph of $La_2CuO_4$ to model, which is close in energy and lattice spacing to the orthorhombic low-temperature phase observed experimentally. The crystal structures were taken from Materials Project [29] and the coordinates were fully optimized. Computed band structure and density of states of bulk FeSe, monolayer FeSe, monolayer FeSe on

SrTiO₃, La₂CuO₄ from this study are shown in the SI. We calculated the paring energy (according to the procedure described in the introduction) and the gap energy in 4 systems based on FeSe, where the Se ion was treated as polarizable. These systems are (i) Bulk FeSe; (ii) one-unit cell thick FeSe film grown on SrTiO₃ (STO) with the oxygen in STO treated as non-polarizable; (iii) one-unit cell of FeSe grown on STO with polarizable oxygen ions; (iv) one-unit cell of FeSe grown on STO with polarizable oxygen ions and one or more monolayers of Se evaporated on the free surface of the FeSe. In order to understand if our model of polarization-induced pairing can be applied and generalized to a different class of compounds we studied one of the simplest cuprate superconductors, doped La₂CuO₄. The cuprates have similarities with the FeSe one-unit cell systems, where they contain highly polarizable oxygen ions and other highly polarizable cations, and the electrical conduction is limited to a narrow 2D region of space centered on the CuO₄ planes.

We then considered the two-charge-carrier Hamiltonian:

$$H_b = H_{a1}(\vec{r}_1) + H_{a2}(\vec{r}_2) + e\emptyset_p(\vec{r}_1,\vec{r}_2) + e\emptyset_R(\vec{r}_1 - \vec{r}_2) \tag{4}$$

The first two terms are the one-charge-carrier Hamiltonians. The third term is the interaction between the two charge-carriers through the potential induced by the probe polarized ions at $\vec{r}_2$. Assuming that the ionic polarizability tensor is constant, independent of the electric field, the reverse process yields the same energy. The fourth term is the Coulomb repulsion between the two electrons.

The electrical potential $\emptyset_p(\vec{r}_1, \vec{r}_2)$ induced by the probe positive charge carrier located at $\vec{r}_1$ was calculated electrostatically. The electric field inducing the polarization in the various ions was calculated by taking screening into account. By assuming the probe is within the FeSe unit cell, the screening is caused by free carriers within the FeSe and is less effective outside. The total electric field induced by the probe both directly and indirectly includes the electric fields induced by all the polarized ions around the probe self consistently. The polarizability we consider is the high frequency one, which involves the displacement of the electrons within the ions. Since the electrons mass is small compared to the nuclei, the polarization map moves with the moving probe with negligible delay.

The Hamiltonian in equation (4) is not periodic in either $r_1$ or $r_2$ but it is periodic in $\vec{r}_1 + \vec{r}_2$ for constant $\vec{r}_1 - \vec{r}_2$. We therefore express the Hamiltonian and wave functions in terms of two new variables:

$$\vec{R} = (\vec{r}_1 + \vec{r}_2)/2 \; ; \qquad \vec{S} = (\vec{r}_1 - \vec{r}_2)/2$$

$$H_b = H_{a1}(\vec{R} + \vec{S}) + H_{a2}(\vec{R} - \vec{S}) + e\emptyset(\vec{R}, \vec{S}) \tag{5}$$

where $e\emptyset(\vec{R}, \vec{S})$ includes the repulsion energy

$$e\emptyset(\vec{R}, \vec{S}) = e\emptyset_p(\vec{R}, \vec{S}) + e\emptyset_R(2\vec{S})$$

We expand the solution in terms of products of two one-charge-carrier Bloch wavefunctions:

$$\Psi_2 = \sum_{\vec{K}, \vec{L}} U_{\vec{K}, \vec{L}} \varphi_{\vec{K}}(\vec{R} + \vec{S}) \, \varphi_{\vec{L}}(\vec{R} - \vec{S}) \exp\left(i \, (\vec{K} + \vec{L}) \cdot \vec{R}\right) \exp(i(\vec{K} - \vec{L}) \cdot \vec{S}) \tag{6}$$

Since the Hamiltonian is periodic in $\vec{R}$, $\vec{K} + \vec{L}$ is a good quantum number. We choose $\vec{K} + \vec{L} = 0$ since we expect as usual that it will lead to the lowest energy. The two-electron Schrodinger equation is then:

$$H_b \psi_2 = E_2 \psi_2 \tag{7}$$

treating $e\emptyset(\vec{R}, \vec{S})$ as a perturbation we solve the equation by solving the matrix equation:

$$\sum_{K', \overline{K > K_F}} [\mathrm{M}(\vec{K}, \vec{K'}) + (2\,\xi_{\vec{K}}) \delta_{K, K'}] \, U_{\vec{K}} = E_2 \tag{8}$$

and $\mathrm{M}(\vec{K}, \vec{K'})$ is the matrix:

$$M(\vec{K}, \vec{K'}) = \iint d^3R\, d^3S\ \varphi^*_{K'}(\vec{R}+\vec{S})\, \varphi^*_{-K'}(\vec{R}-\vec{S})\, e\emptyset(\vec{R},\vec{S})$$

$$\varphi_{-K}(\vec{R}-\vec{S})\, \varphi_K(\vec{R}+\vec{S})\, \exp(i2(\vec{K}-\vec{K'})\cdot\vec{S}) \tag{9}$$

We evaluated the periodic part of the Bloch states lying ~ 2 eV above and below the Fermi level on a real-space grid, and calculated the matrix elements of Eq. (9) on a regular mesh of k-points. $\vec{K}, \vec{K'}$ were both outside the Fermi surface and the corresponding one-electron energies were less than 0.5 eV above the Fermi energy. $\vec{r_1}$ was limited to one unit-cell because of the periodicity and $\vec{r_2}$ was limited to the region, where $\emptyset(\vec{R},\vec{S})$ is significant, which yielded an (N x N) matrix. Solving equation (8) yields N solutions. The solution with the lowest total energy, i.e. $\sum_{\overline{K>K_F}}[2\,\xi_{\vec{K}}\delta_{K,K'}]\, U_{\vec{K}} + E_2$, is the energy of the paired electrons provided that the total energy is negative.

The energy of the multi-electron BCS wave function is expressed in terms of the energies $\Delta_{\vec{K}} \geq 0$. Detailed discussion can be found in previous work [9] [27].

$$\Xi = \sum_{\vec{K}} \xi_{\vec{K}}\left(1 - \frac{\xi_{\vec{K}}}{E_{\vec{K}}}\right) + \frac{1}{4}\sum_{\vec{K},\vec{K'}} M(\vec{K},\vec{K'}) \frac{\Delta_{\vec{K}}\Delta_{\vec{K'}}}{E_{\vec{K}}E_{\vec{K'}}} \tag{10}$$

Here

$$E_{\vec{K}} = \sqrt{\xi_{\vec{K}}^2 + \Delta_{\vec{K}}^2}$$

If all $\Delta_{\vec{K}} = 0$, then $\Xi = 0$, and the BCS wavefunction corresponds to a superconducting state if $\Xi < 0$, which can happen only if $M(\vec{K}, \vec{K'}) < 0$ for some values of $\vec{K}, \vec{K'}$. We searched for 3 different $\vec{K}$ values such that the sum of all their contributions to Eq. (10) was the most negative. We set all other $\Delta$ to zero. If the value of $\Xi$ is negative, this value is the closest to zero from below. In principle with proper choice of $\Delta_{\vec{K}}$ a lower energy can be found.

**Results and discussion**

One central result of this paper is that the polarizability model accounts correctly and quantitatively for the behavior of three materials belonging to three different groups:

crystalline FeSe (system i); one monolayer FeSe grown on STO with polarizable Oxygen (system (iii); and doped $La_2CuO_4$. In addition, by comparing the results in systems ii and iii (Oxygen non-polarizable and polarizable) we found that the polarizability of the oxygen in the top layers of STO plays a very important role in driving the FeSe film to superconductivity. In addition, we predict that evaporating Se on the free surface of the FeSe film (system iv) will drive up the paring energy.

As discussed in the introduction the potential induced by the probe plays a decisive role in pairing the charge carriers. In Fig. 1 we present five examples of potential maps of FeSe-based systems and $La_2CuO_4$. These potential maps were obtained with the probe at positions specified in the figure caption. The potential function $\emptyset(\vec{R},\vec{S})$, which plays a key role in the interaction between the two charge carriers, was calculated with the probe at positions on a grid within the central unit cell and the potential on a grid involving 7x7x1 unit cells (7x7x5 for bulk FeSe). While the full potential matrixes were used in the calculations, to highlight the negative potential areas, we artificially collapsed in the figure, the positive potentials above 0.1 to 0.1 eV and the negative potentials below -0.3 to -0.3 eV. Fig. 1(a) shows a slice of the potential map of FeSe on STO treating the oxygen ions as non-polarizable, while Fig. 1(b,c) are for bulk FeSe, where we can notice that the area and size of the negative (i.e. attractive) potential are small for both cases. In contrast, the potential maps of one-unit-cell of FeSe grown on STO with polarizable oxygen (system iii) Fig.(1d) and with Se evaporated on FeSe free surface (system iv) Fig.(1e) revealed negative potential large both in size and area contributing to large paring and gap energies. Fig. (1f) shows an example of the potential map of $La_2CuO_4$. Here again the potential size and area are large.

As seen in Fig.1 the negative potential extends to about 3-unit cells so the second charge carrier will be roughly located within this region with a small probability to be found outside. This is typical of most high Tc superconductors.

The total pairing energy for the four systems as a function of polarizability at the Fermi level 0.4 eV is shown in Fig. 2. The polarizability scale is that of the Se and the polarizability of the oxygen is equal to ~5/8 of selenium [15]. Our calculations show that bulk FeSe is not superconducting as indicated by the observation that the pairing energy is very small but positive, independent of the Selenium/Oxygen polarizability. This is in reasonable agreement with the experimentally observed $T_c$ of ~9 K for bulk

FeSe [30]. The difference can be attributed to small contributions of other mechanisms such as electron-phonon interaction and spin-fluctuations. In contrast, our results show that monolayer FeSe on STO becomes superconducting with polarizability as large as $\alpha(Se)>3.2$ and $\alpha(O)>0.8$ in units of $10^{-40}$ $Cm^2/V$. Since the interaction among charge carriers is strong, pairing energy needed for the observed $T_c$ may be as large as 5 $k_B$ $T_c$. To account for $T_c=100$ K we need a paring energy $E_p$ of at least 0.042 eV, according to Fig. 2, which corresponds to anion polarizabilities of $\alpha(Se)=3.4$ $10^{-40}$ $Cm^2/V$ and $\alpha(O)>2.2$ $10^{-40}$ $Cm^2/V$. These values are well within the range reported in literature [15, 16]. Furthermore, our results suggest that evaporating Se on the free FeSe surface would drive the pairing energy vs polarizability curve further up (Figure 2). The known polarizability of neutral atomic Se is between 4.29 and 4.785 $10^{-40}$ $Cm^2/V$ [31]. In our calculations, we assumed that the evaporated Se fully covers the FeSe surface, and in practice the surface may not be fully covered and the Se film will be disordered, which may decrease the pairing energy.

We further report the pairing energy and gap energy of one-unit-cell-thick FeSe on STO as a function of the Fermi level for fixed values of the polarizability. The Fermi level can play an important role since it determines the free charge density and therefore the screening. While the screening can decrease repulsion inside the film, thus increasing the pairing energy, it can decrease the electric field that polarizes the ions, thus decreasing the pairing energy. The fact that the charge carriers are free to move only in the FeSe film means that the screening within the film is more effective than perpendicular to it, which increases the pairing energy as long as the screening is weak but decreases the pairing energy when the screening is strong. Fig. 3 shows that the pairing energy becomes more negative with increasing the Fermi energy for low Fermi energy values because repulsion is large, while the pairing energy becomes more positive with increasing Fermi energy for larger Fermi energy values because the electric field inducing the polarization is smaller. In addition, the gap energy shown in Fig. 3 is smaller than the pairing energy but both quantities follows similar trends. The pairing and gap energies dependence on the Fermi level is qualitatively consistent with the volcano-shaped $T_c$ dependence on doping [32]. However, one should bear in mind that doping affects not only screening, but also the polarizabilities, which should be considered when comparing the doping dependence of pairing and gap energies with experiments.

Our computation results and model explain the behavior of the oxygen isotope effect. The free charge carrier spends about $10^{-15}$ sec in one unit-cell. During this time the electrons in the ions can respond but not the heavy cores. Consequently, the oxygen isotope effect is barely detectable. The value we obtain for the gap energy is about 9 meV, in good agreement with the value that Song et. al. [24] identify as the contribution from another channel.

To test the applicability of the polarizability model to other types of superconductors we show the pairing and gap energies of $La_2CuO_4$ (at the Fermi level $E_F = 0.25$ eV) as a function of average polarizability in Fig. 4. We observe that the energy of the two-electron wavefunction of this system becomes negative, namely the electrons become paired, with both α(La) and α(O) > 2.5 $10^{-40}$ $Cm^2/V$. The experimental $T_c = 40$ K of $La_2CuO_4$, which corresponds to a pairing energy of $E_p \sim 5\ K_BT_c = 0.017$ eV, requires α = 2.7 $10^{-40}$ $Cm^2/V$ for O and La. These values are well within the range reported in literature [15,16]. The pairing and gap energies at polarizability α = 3 $10^{-40}$ $Cm^2/V$ for $La_2CuO_4$ as a function of the Fermi level are shown in Fig. 5. The observed behavior is similar to that of FeSe monolayer on STO with polarizable oxygen ions (Fig. 4). The pairing and gap energies approach zero at small Fermi level because the screening of the repulsive field is not effective enough while at large Fermi level the screening of the polarizing electric field is too large and superconductivity is suppressed.

**Summary and conclusions**

High $T_c$ superconductors possess in general large electronic polarizabilities. We have shown that under proper conditions these polarizabilities may lead to electron pairing and superconductivity. We find that two of the systems we investigated, namely one-unit cell film of FeSe on STO and doped $La_2CuO_4$, are superconductors. We find that above a certain level of polarizability the lowest two-charge-carrier wave-function energy is negative namely below Fermi energy, meaning that the two charge carriers are paired. Moreover, with polarizability of about 4 $10^{-40}$ $Cm^2/V$ and 2.5 $10^{-40}$ $Cm^2/V$ for Se and O, respectively, and 2.7 $10^{-40}$ $Cm^2/V$ for La and O for $La_2CuO_4$, the paring energy is 5 $K_BT_c$. These polarizability levels are well within the range of values reported in the literature. In conclusion, we have shown that the electronic polarizability of oxygen and other highly polarizable ions contribute decisively to charge carrier pairing and to superconductivity in high $T_c$ superconductors. The polarizabilities needed to achieve the experimentally established $T_c$, pairing energy and gap energy are within the range of values determined theoretically and experimentally. Moreover, our results show that the pair distance is small, less than about 1 nm, for the materials studied in this work, in agreement with experiment [33].

The fact that the electronic polarizability model explains correctly and quantitatively the super conductivity parameters, gap and paring energies, of two types of high Tc materials pnictides and cuprates with similar polarizability parameters suggests that the same model may be applicable to other material systems within these groups as well as other high Tc materials. It should be emphasized that one has to consider explicitly both the polarizability of the ions and the actual structure of the material. It is also important to test the model predictions with respect to various parameters experimentally, for example, by measuring the paring energy and Tc of one-unit cell of FeSe on STO with Se evaporated on the free FeSe surface. In addition, the polarizability model may provide a tool for identifying new high Tc materials and to test its predictions quantitatively.

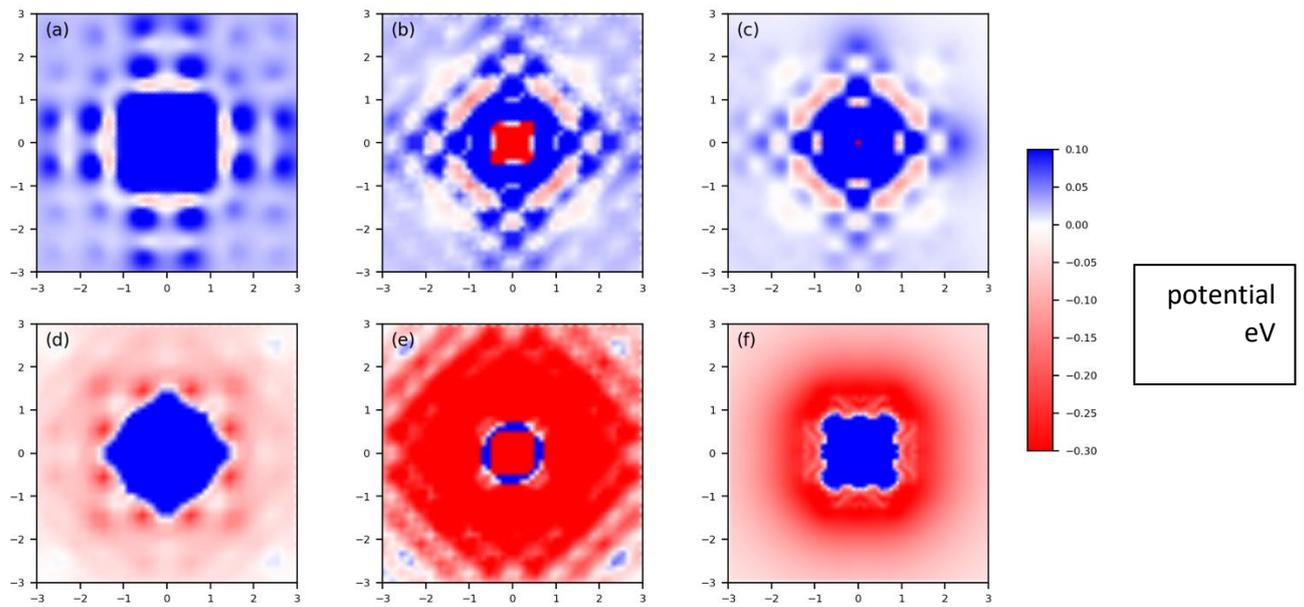

FIGURE 1. Examples of electric potential maps as a function of position in one-unit cell units. The potential maps are induced by a charge carrier (the probe) located at the center of a FeSe unit cell as indicated in supplementary figure S1s by the arrow pointing to [x=0; y=0; z=0]. The potential is shown on planes inside and parallel to the FeSe film at the heights indicated below. The corresponding plane height in $La_2CuO_4$ is relative to the $CuO_4$ plane. Maps b and e show attractive potential at the center because these planes are above and below the probe. The potential is calculated for probes occupying 8x8x6 points within one-unit cell. The potential is calculated in 8x8x1 unit cells for 8x8x6 points within one cell. The unit of length is one-unit cell. The pairing potential is in eV.

a) z = 0; One-unit cell thick FeSe film on $SrTiO_3$ with non-polarizable oxygen.
b) z = -0.25; Bulk FeSe
c) z = 0; Bulk FeSe
d) z = 0; One-unit cell thick FeSe film on $SrTiO_3$ with polarizable oxygen
e) z = 0.25; Same with Se evaporated on the FeSe free surface
f) z = 0; $La_2CuO_4$. The probe is located in the center of the $CuO_4$ cell.

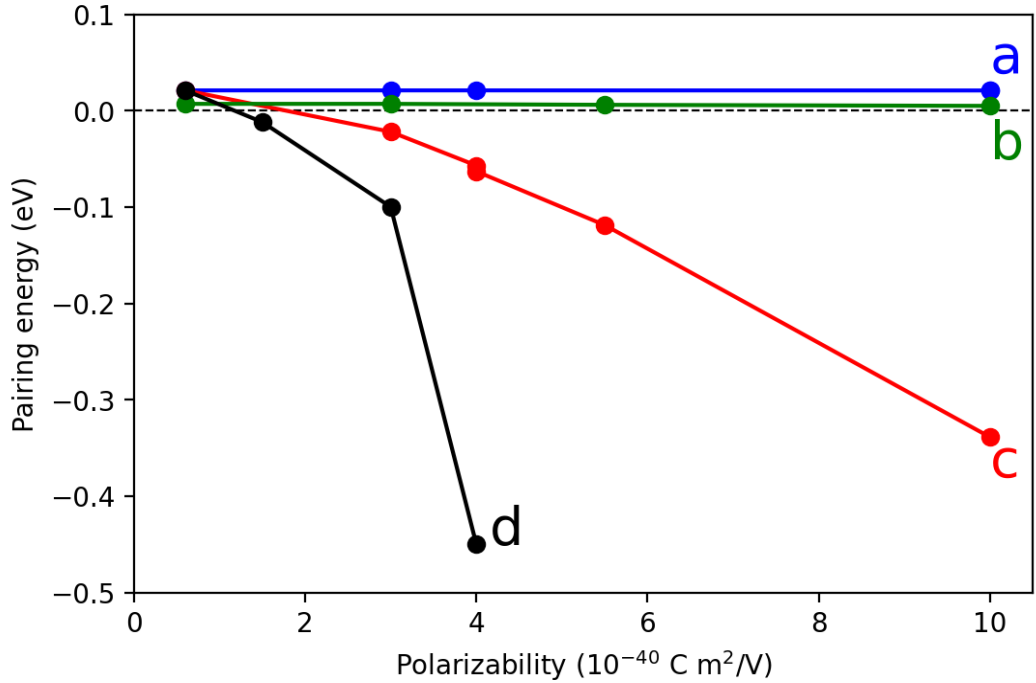

FIGURE 2. Pairing energy vs. Se and O polarizability. (a) 1-unit cell of FeSe on $SrTiO_3$. Here selenium is polarizable but the oxygen is assumed to be non-polarizable. (b) Bulk FeSe, where Se is polarizable. (c) 1-unit-cell of FeSe on $SrTiO_3$. Both oxygen and selenium are polarizable. (d) 1-unit cell of FeSe on $SrTiO_3$ with Se evaporated on the free surface of the FeSe film. Both O and Se are polarizable. The polarizability on the x axis is the one of Se, and for the cases where oxygen is polarizable it is assumed that the polarizability of O is equal to 5/8 of Se.

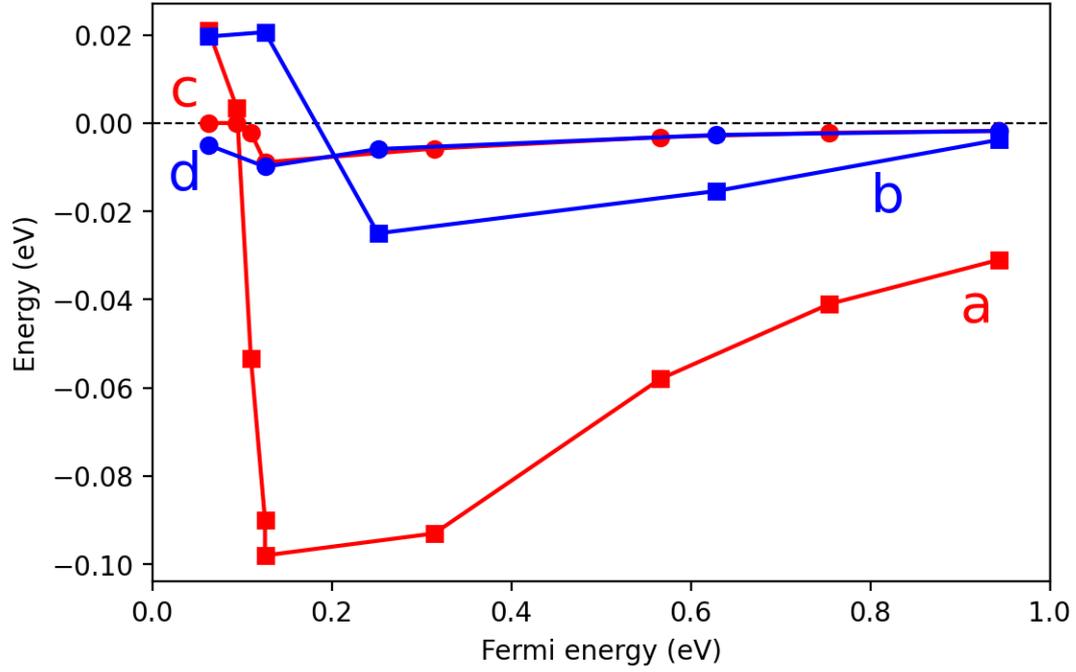

FIGURE 3. Pairing and gap energy vs. the Fermi energy for constant polarizabilities in one FeSe unit cell on $SrTiO_3$. (a) Pairing energy for Se and O polarizabilities of 5.0 and 3.125 $10^{-40}$ C m$^2$/V. (b) Pairing energy for Se and O polarizabilities of 3.0 and 2.125 $10^{-40}$ C m$^2$/V. (c) Gap energy for Se and O polarizabilities of 5.0 and 3.125 $10^{-40}$ C m$^2$/V. (d) Gap energy for Se and O polarizabilities of 3.0 and 2.125 $10^{-40}$ C m$^2$/V.

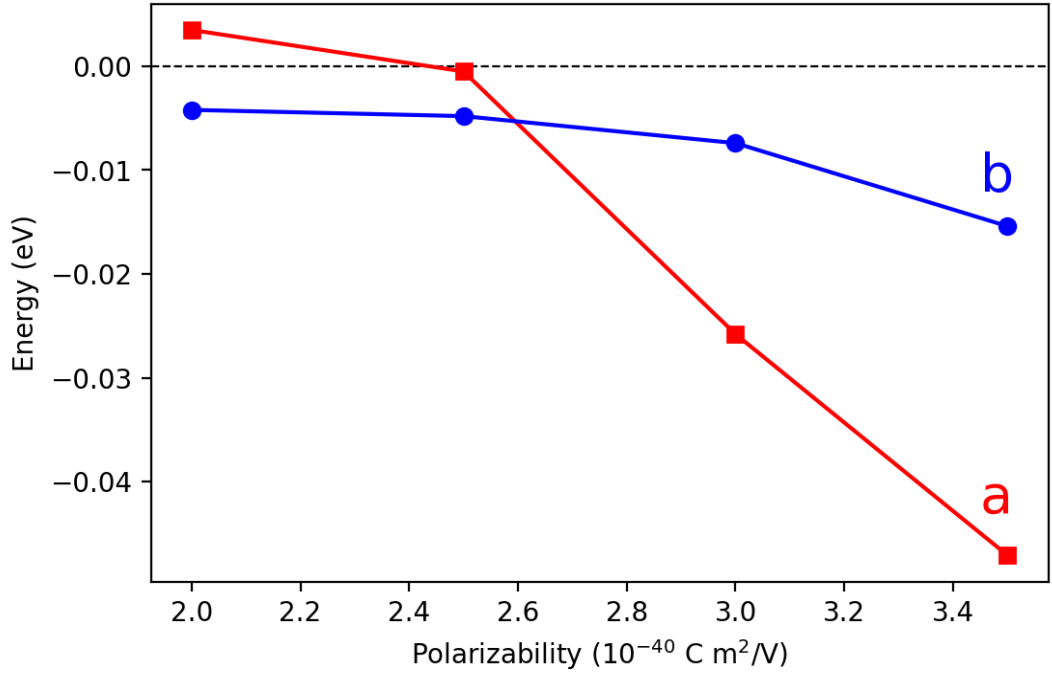

FIGURE 4. Pairing and gap energy in $La_2CuO_4$ as a function of ionic polarizability for constant Fermi energy equal to 0.25 eV. (a) Pairing energy in eV. (b) Gap energy in eV.

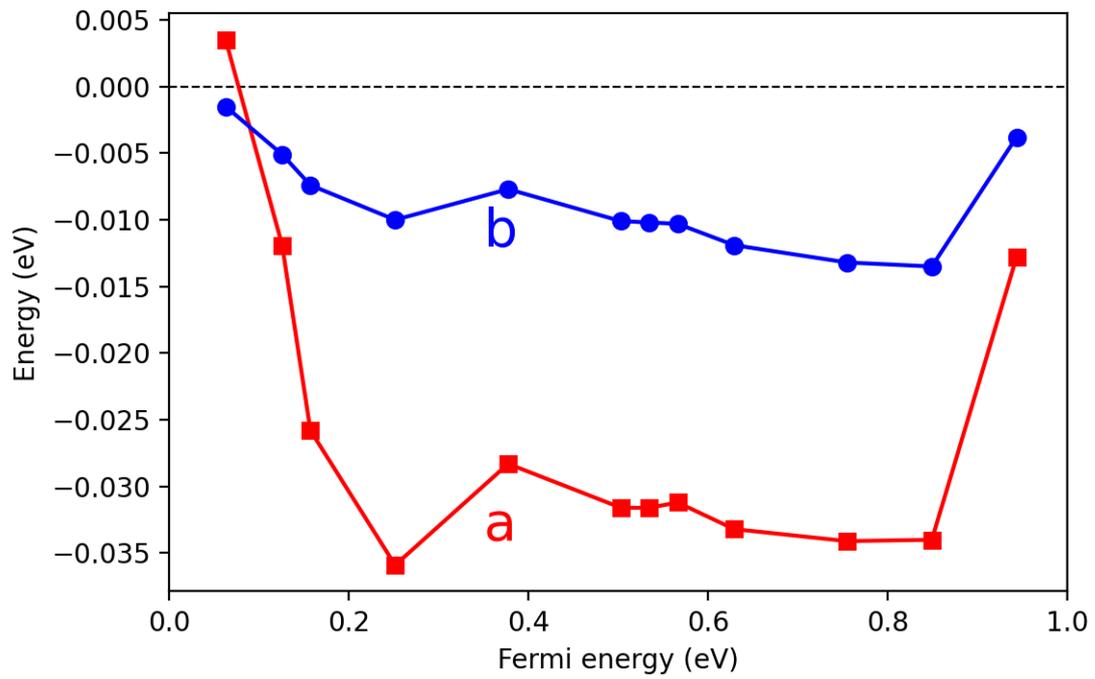

FIGURE 5. Pairing and gap energy in $La_2CuO_4$ as a function of Fermi energy for constant O polarizability of $3 \times 10^{-40}$ C m$^2$/V. (a) Pairing energy in eV. (b) Gap energy in eV.

# Supplementary information for:

# Electronic Polarizability Induced Cooper-like Pairing and Energy Gap in High-$T_c$ superconductors

Yizhak Yacoby [1], Davide Ceresol[2], Livia Giordano[3,4,5] and Yang Shao-Horn[3,4,6,7]

1. Racah Institute of physics, Hebrew University, 91904, Jerusalem, Israel
2. Consiglio Nazionale delle Ricerche, Istituto di Scienze e Tecnologie Chimiche "G. Natta" (CNR-SCITEC), 20133 Milan, Italy
3. Electrochemical Energy Laboratory, Massachusetts Institute of Technology, Cambridge MA 02139, USA
4. Research Laboratory of Electronics, Massachusetts Institute of Technology, Cambridge MA 02139, USA
5. Dipartimento di Scienza dei Materiali, Università di Milano-Bicocca, Milano, Italy
6. Department of Mechanical Engineering, Massachusetts Institute of Technology, Cambridge MA 02139, USA
7. Department of Materials Science and Engineering, Massachusetts Institute of Technology, Cambridge MA 02139, USA

**1. Details of DFT calculations**

We performed DFT calculations with the planewave pseudopotential code Quantum Espresso [1]. We used norm-conserving pseudopotentials from the ONCV set [2] and a planewave cutoff of 90 Ry.
In the case of $La_2CuO_4$ we use the rotationally invariant DFT+U method [3] with a Hubbard U of 4 eV on the Cu ions in order to open a band gap of similar magnitude to that reported in Ref. [4]. All calculations were not spin-polarized except $La_2CuO_4$ which was calculated in the antiferromagnetic (AFM) state.
The periodic part of the Bloch wavefunctions were calculated on a regular grid in the full Brillouin zone and exported to MATLAB on a real space grid using *wfck2r.x* code of the QE package.

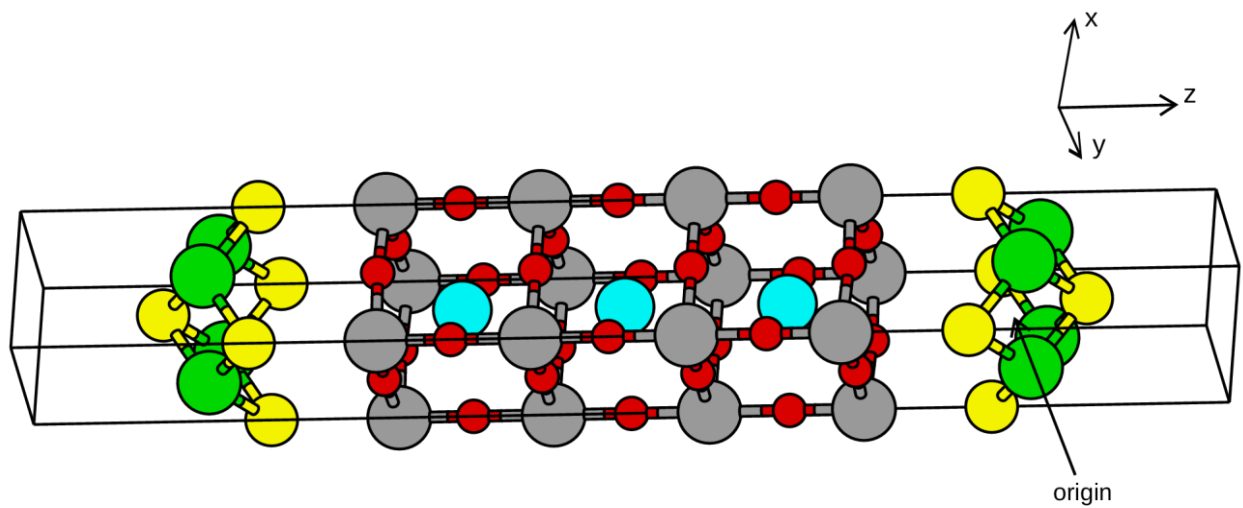

FIGURE S1. Atomic structure of FeSe (Se-yellow, Fe-green) on 7-layer $TiO_2$-terminated $SrTiO_3$ (Sr-blue, Ti-gray, O-red). The arrow indicates the origin of the probe charge.

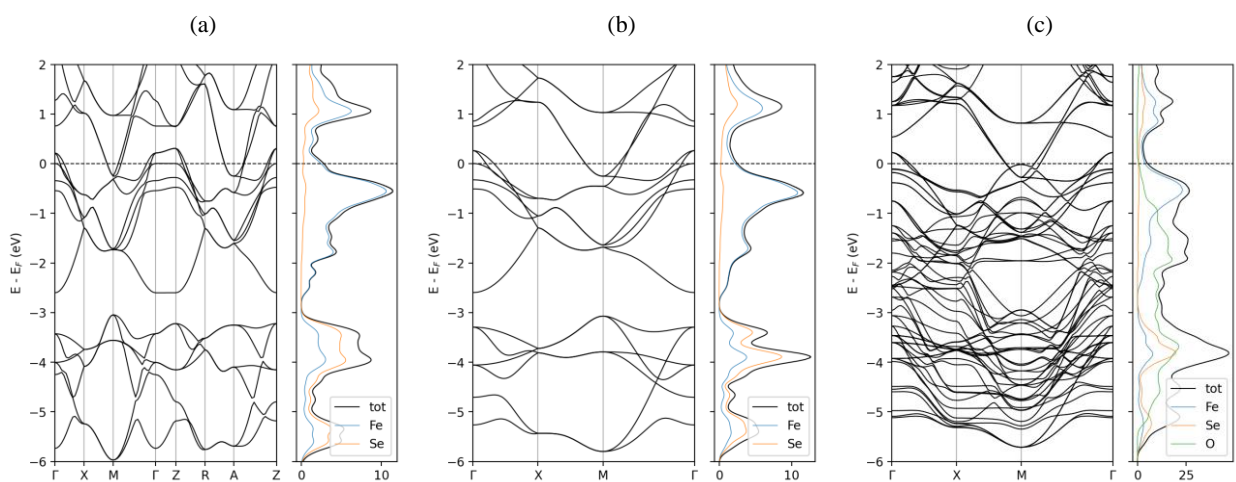

FIGURE S2. Band structure and projected density of states of (a) bulk FeSe; (b) free standing FeSe monolayer; (c) FeSe monolayer on $SrTiO_3$. Note that the band structure of bulk FeSe along the Γ-X-M-Γ path is very similar to that of monolayer FeSe. In both cases that are two hole pockects around the Γ point and one electron pocket around the M point. The states at the Fermi level have a large Fe *3d* character. The band structure of FeSe on $SrTiO_3$ also shows a hole pocket at G and an electron pocket at M. The oxygen orbitals contribute mostly to the states below –1 eV, and with the appearance of an extra band touching Fermi level from below at the M point.

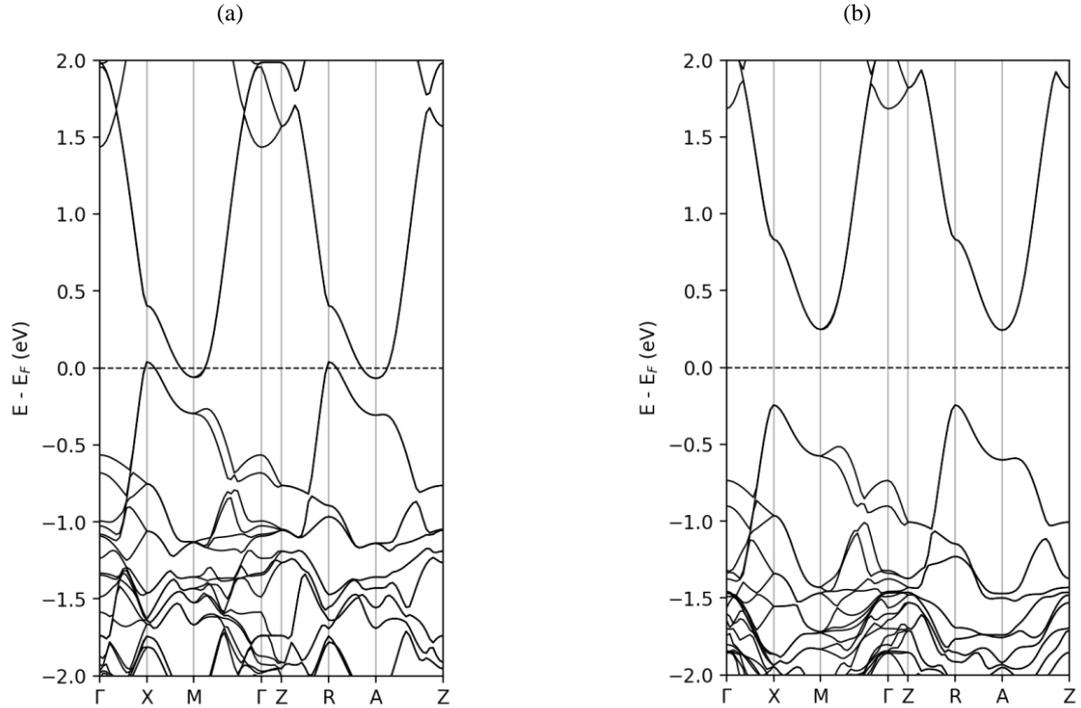

FIGURE S3. Band structure of the low temperature tetragonal (LTT) phase of $La_2CuO_4$ in the AFM state; (a) DFT; (b) DFT+U. Note that without Hubbard U correction, the system is predicted to be a metal.